\documentstyle[eqsecnum,epsfig,aps,prd]{revtex}
\let\jnfont =\rm
\def\NPB#1,{{\jnfont Nucl.\ Phys.\ B }{\bf #1},}
\def\PLB#1,{{\jnfont Phys.\ Lett.\ B }{\bf #1},}
\def\PRD#1,{{\jnfont Phys.\ Rev.\ D } {\bf #1},}
\def\PRL#1,{{\jnfont Phys.\ Rev.\ Lett.\ }{\bf #1},}
\def\ZPC#1,{{\jnfont Z.~Phys.\ C }{\bf #1},}
\begin{document}
%\draft
%\preprint{ VLBL Study Group-H2B-6\\ AMES-HET-02-05 }

\hfill\vtop{
\hbox{VLBL Study Group-H2B-6}
\hbox{AMES-HET-02-05}
\hbox{hep-ph/0208193}
\hbox{}}

\vspace*{.25in}
\begin{center}
{\large\bf Measuring $CP$ violation and mass ordering in\\
joint long baseline experiments with superbeams}\\[10mm]
K. Whisnant$^a$, Jin Min Yang$^b$, Bing-Lin Young$^a$\\[5mm]\it
$^a$ Department of Physics and Astronomy, Iowa State
University, Ames, Iowa 50011, USA\\
$^b$ Institute of Theoretical Physics, Academia Sinica, Beijing 100080, China
\end{center}
\date{\today}
\thispagestyle{empty}

\begin{abstract} 

We propose to measure the $CP$ phase $\delta_{\rm CP}$, the magnitude
of the neutrino mixing matrix element $|U_{e3}|$ and the sign of the
atmopheric scale mass--squared difference $\Delta{\rm m}^2_{31}$ with
a superbeam by the joint analysis of two different long baseline
neutrino oscillation experiments. One is a long baseline experiment
(LBL) at 300 km and the other is a very long baseline (VLBL)
experiment at 2100 km.  We take the neutrino source to be the approved
high intensity proton synchrotron, HIPA.  The neutrino beam for the
LBL is the 2-degree off-axis superbeam and for the VLBL, a narrow band
superbeam. Taking into account all possible errors, we evaluate the
event rates required and the sensitivities that can be attained for
the determination of $\delta_{\rm CP}$ and the sign of $\Delta
m^2_{31}$.  We arrive at a representative scenario for a reasonably
precise probe of this part of the neutrino physics.

\end{abstract}  

\section{Introduction}
\label{sec1}

The Super-Kamiokande experiments~\cite{superK} in the past several
years, joined by SNO~\cite{other} more recently, have given strong
indications of neutrino oscillation that are corroborated and
constrained by a variety of other experiments. These experiments
started a new era in the study of neutrino physics and offered the
best indication to date of physics beyond the standard model.  To
further probe neutrino physics, there are a number of ongoing and
planned neutrino oscillation experiments.  These experiments promise
to give a full description of the phenomenology of neutrino mixing.
The most attractive experiments among the new generation of neutrino
oscillation experiments are the long baseline (LBL) experiments.  They
are performed in the controlled environment of traditional
experimental high energy physics and expected to allow precision
measurements of the oscillation parameters, including the leptonic
$CP$ phase.  Notably, the recently approved superbeam facility, the
High Intensity Proton Accelerator (HIPA)~\cite{HIPA}, which can
provide intensive high energy neutrino beams from its 50 GeV proton
synchrotron, offers the possibility of even more desirable LBL
experiments.  So far, the possibility of two LBL experiments using the
HIPA superbeam have been discussed.  One is HIPA to Kamiokande at a
baseline length of about 300~km ~\cite{J2K} known as J2K, and the
other is HIPA to a detector located 2100 km away near
Beijing~\cite{H2B,Japanesegroup} called H2B. It is well-known that
there are parameter ambiguities that are generally associated with
oscillation measurements at a single baseline\cite{ambiguities,huber}.
Measurements at more than one baseline can be
beneficial~\cite{Japanesegroup,bmw2001,mena}; our previous
studies~\cite{joint,optimum} showed that the joint analysis of the J2K
and H2B experiments can offer extra leverages to resolve some of these
ambiguities. Our results, however, also showed that $CP$ violation
effects cannot be determined at 3$\sigma$ level even with the joint
analysis considered in the study, in which no antineutrino beams were
used.

Since the leptonic $CP$ phase and mass--squared difference sign are
pertinent information in the physics of neutrino mixings, which seems
to be very different from that of the quark sector, it is necessary to
find out how to pin down these neutrino mixing parameters accurately
in new experiments.  It has been widely recognized that the neutrino
factory~\cite{geer} is an ideal facility for the study of neutrino mixings.
However, because of the technical and budgetary challenges faced with
building a workable neutrino factory in the near future, and because
of the availability of a conventional superbeam from HIPA in about
five years, it is obviously advisable that we explore the full
potential of the HIPA superbeams.  In this work, we examine in further
detail the measurement of the $CP$ phase and the mass--squared
difference sign from the joint analysis at the two long baselines with
specific HIPA superbeams and more suitable detectors.  The neutrino
beams are the 2-degree off-axis superbeams for the LBL at 300 km and a
narrow band superbeam for VLBL at 2100 km. We evaluate the event rates
and investigate their sensitivities to the $CP$ phase and the sign of
$\Delta m^2_{31}$.  Taking into account all possible experimental
errors, we find that a fairly precise measurement of the $CP$ phase, the
sign of the mass--squared difference and the mixing angle
$\theta_{13}$ is possible but requires: (1) the joint analysis at the
two baselines; (2) that both a $\nu_{\mu}$ beam and a $\bar \nu_{\mu}$
beam are needed at 300 km; (3) that a $\bar \nu_{\mu}$ beam is needed
at 2100 km if $\Delta m^2_{31}$ is negative; and (4) a significant
increase in the statistics at both 300 km and 2100 km.  We find that
$\sin^2(2\theta_{13})$ can be probed to very small values, depending
on the value of the $CP$ phase.

In Sec.~II we describe how the simulations are performed.  In Sec.~III
our results are presented.  A brief discussion and conclusion can be
found in Sec.~IV.

\section{Description of simulations }
\label{sec2}
\subsection{Parametrization and inputs}

Our oscillation analyses will be restricted to 3 flavors of active
neutrinos.
%The oscillation of  the 3-flavor neutrinos is a system with a limited number 
%of degrees of freedom. 
The parameters of the system consists of 2 mass--squared differences
(MSD), 3 mixing angles and 1 measurable $CP$ phase.  The unitary
mixing matrix in the vacuum is parameterized as usual
\begin{eqnarray} 
U & = & \left( \begin{array}{ccc} c_{12}c_{13} & c_{13}s_{12} &
    \hat{s}^*_{13} \\ -c_{23}s_{12} - c_{12}\hat{s}_{13}s_{23} &
    c_{12}c_{23} -s_{12}\hat{s}_{13}s_{23} & c_{13}s_{23} \\
    s_{12}s_{23} - c_{12}c_{23}\hat{s}_{13} & -c_{12}s_{23}
    -c_{23}s_{12}\hat{s}_{13} & c_{13}c_{23} \end{array} \right),
\end{eqnarray}
where $s_{jk}=\sin(\theta_{jk})$, $c_{jk}=\cos(\theta_{jk})$, and
$\hat{s}_{jk}=\sin(\theta_{jk})e^{i\delta_{\rm CP}}$, $\theta_{jk}$
defined for $j<k$ are the mixing angles of mass eigenstates $\nu_j$
and $\nu_k$, and $\delta_{\rm CP}$ is the $CP$ phase angle.  The three
mass eigenvalues are denoted as $m_1$, $m_2$, and $m_3$.  The two
independent MSD are $\Delta{\rm m}^2_{21}\equiv {\rm m}^2_2 - {\rm
m}^2_1$ and $\Delta{\rm m}^2_{31}\equiv {\rm m}^2_3 - {\rm m}^2_1$.

The inputs of the mixing angles and MSD's are obtained from solar,
atmospheric and reactor experiments:
\begin{eqnarray} 
& \sin^2(2\theta_{12})=0.8,& ~~~ \Delta{\rm m}^2_{21}=5 \times 10^{-5}
                 {\rm eV}^2,
\label{input12} 
\\ & \sin^2(2\theta_{23})=1.0,& ~~~ |\Delta{\rm m}^2_{31}|=3 \times
10^{-3} {\rm eV}^2,
\label{inpu23}
\\ & \sin^2(2\theta_{13}) \leq 0.1. &
\label{input13}
\end{eqnarray}
Note that the sign of $\Delta m^2_{31}$ is unknown. The currently
favored Large Mixing Angle solar solution requires $\delta m^2_{21} >
0$.

In LBL experiments the neutrino beam interacts with electrons
contained in Earth matter~\cite{matter}.  In the present study we use the
Preliminary Reference Earth Model~\cite{earth}, generally known as
PREM, for Earth density profiles~\cite{profile} and numerically
integrate the Schr\"{o}dinger equation that descibes the propagation
of the neutrino in matter for the treatment of distance dependent
matter density.  However, we note that there exist more sophisticated
approaches to Earth matter effect, including both an updated average
density profile known as the AK135~\cite{AK135} and treatments of
uncertainties of the density profile~\cite{huber,shan,matteruncertainty}
which can affected the determination of the $CP$ phase angle.

\subsection{Beams and detectors}

The HIPA 50 GeV proton synchrotron beam calls for a power of 0.77 MW
in phase I, to be upgraded to 4 MW in phase II. The superbeam provided
by HIPA can be a wide band beam (WBB), a pulsed narrow band beam
(NBB), or an off-axis beam (OAB).  The WBB contains neutrinos with
widely distributed energy.  In a NBB the neutrino flux is concentrated
in a narrow range of energies, with maximum energy $E_{\rm peak}$
where the intensity is peaked, and the intensity decreases rapidly
below $E_{\rm peak}$.  An OAB also peaks at a certain energy, but has
a longer high-energy tail than that of a NBB.  More details of the
various beam profiles can be found in Ref.~\cite{oab2}.  For the 2100
km baseline we use a NBB~\cite{beam} with peak energy of 4 GeV.  We
have also investigated the beams of peak energies of 5, 6 and 8 GeV,
but found that the 4 GeV beam gives the best results.  For 300 km we
use the 2-degree OAB (2$^\circ$--OAB) which has a peak energy at 0.8
GeV~\cite{oab}.

The detector at 2100 km is assumed to be a water Cerenkov calorimeter
with resistive plate chambers~\cite{wang,H2B} located in Beijing,
tentatively called the Beijing Astrophysics and Neutrino Detector
(BAND).  The size of the detector will be 100 kt at the beginning and
can be upgraded to a much larger one depending on the physics
requirements.  The detector at 300 km is initially the Kamiokande
detector of the present size of 22.5 kt and upgraded later to 450
kt.\footnote{For a more discussion of the 300 km detector we refer to
Ref.~\cite{J2K}.}

\subsection{Experimental errors}

For the experimental errors we use 3$\sigma$ throughout this work.
All independent errors, statistical and systematic, are added in
quadrature.  For the statistical error we used $3\sigma$ Poisson
errors as described in the appendix of Ref.~\cite{bgrw}, including a
background at the 1\% level of the rate of the survival channel. For
the systematic error we assumed that the background is known at the
2\% level as given in ~\cite{jhfsk}.  To estimate the error due to the
uncertainty in the measurement of the mixing angle $\theta_{23}$, we
assumed that $\sin^2(2\theta_{23})$ is measured via the survival
channel at $L = 300$~km, with the event rate given by $N(\nu_\mu \to
\nu_\mu) \simeq N_0 ( 1 - \sin^2(2\theta_{23}) \sin^2(\Delta))$, where
$\Delta \simeq \pi/2$ and $N_0$ is the number of events in the absence
of oscillations. Then the statistical uncertainty on
$\sin^2(2\theta_{23})$ is
\begin{equation}
\delta(\sin^2(2\theta_{23})) = \sqrt{N}/N_0 \approx {1\over
                 \sqrt{N_0}}\sqrt{( 1 - \sin^2(2\theta_{23})
                 \sin^2(\Delta))}\,.
\label{eq:uncertainty}
\end{equation}
We then find the variation in the rate in question for a 3$\sigma$
deviation in $\theta_{23}$, and added this in quadrature to the other
3$\sigma$ errors described above to obtain the total 3$\sigma$ error.
 
\subsection{Scenarios}

We consider three scenarios in the present investigation.  The
scenarios are summarized in Table~1.
%\ref{tab:scenariotable} 
For Scenario I the first stage involves a 5-year experiment with a
water Cerenkov detector of 22.5 kt at 300 km with the 2$^\circ$--OAB
$\nu_{\mu}$ beam~\cite{oab} from HIPA at 0.77 MW.  This stage is
contained in the plan of J2K~\cite{J2K}.  The second stage of this
scenario has HIPA upgraded to 4 MW, as discussed in Ref.~\cite{H2B},
to deliver a NBB $\nu_{\mu}$ beam to the water Cerenkov detector of
100 kt at L=2100 km to run for 5 years.

Scenario II has an upgraded 4 MW HIPA and calls for both
$\nu_{\mu}$ and $\bar \nu_{\mu}$ beams.  The experiment at 2100 km is
the same as in Scenario I.  For 300 km, however, we assume a much
larger water Cerenkov detector of 450 kt. It will run for 2 years with
2$^\circ$--OAB $\nu_{\mu}$ beam, and then 6 years with the 2$^\circ$--OAB $\bar
\nu_{\mu}$ beam.

Scenario III is similar to Scenario II, but calls for a much larger
water Cerenkov detector at 2100 km, to run either the $\nu_\mu$ or
$\bar\nu_\mu$ beam, for example, for 5000 kt-yr.  Whether a $\nu$ or
$\bar \nu_{\mu}$ beam is delivered to the 2100~km site depends on the
sign of $\Delta{\rm m}^2_{31}$, as will be explained below.
%%%%%%%%%%%%%%%%%%%%%%%%
%\null\vspace{0.2cm}
%\noindent
%{\small Table 1: Scenarios in our  investigation.}
\vspace{0.03in}

%\begin{table}[hbtp!]
%\begin{center}
%\begin{tabular}{|l|c|c|c|c|c|c|}  %\hline   
%           & \multicolumn{3}{c|}{ } & \multicolumn{3}{c|}{ } \\ 
%           & \multicolumn{3}{c|}{L=300 km} & \multicolumn{3}{c|}{L=2100 km} \\ 
%           & \multicolumn{3}{c|}{ } & \multicolumn{3}{c|}{ } \\  \cline{2-7}
%           & & & & & & \\ 
%           & beam & power & detector size $\times$  runing time  & beam & power
%                                      & detector size $\times$ runing time \\  
%           & (2$^\circ$--OAB)& (MW)  & (kt$\cdot$ year)  & (NBB)&(MW) & (kt$\cdot$year)       %\\ 
%           & & & & & &  \\ \hline \hline
%           & & & & & & \\
%Scenario I & $\nu_{\mu}$ &  $0.77$ & $22.5\times 5$ & $\nu_{\mu}$ & $4$  
%                & $100\times 5$  \\  
%           & & & & & & \\ \hline \hline
%           & & & & & &  \\
%Scenario II  & $\nu_{\mu}$&    $4$    & $450\times  2$    & $\nu_{\mu}$& $4$  
%                  & $100\times 5$  \\ 
%          & & & & & & \\ 
%            & $\bar \nu_{\mu}$& $4$    & $450\times  6$       &            &      &  \\ 
%           & & & & & &  \\ \hline \hline
%           & & & & & &  \\
%Scenario III & $\nu_{\mu}$    & $4$    & $450\times  2$  & $\nu_{\mu}$     & $4$ 
%          & $5000$ for $\Delta{\rm m}^2_{31}>0$ \\ 
%           & & & & & &  \\
%          & $\bar \nu_{\mu}$& $4$    & $450\times  6$  & $\bar \nu_{\mu}$  & $4$ 
%           & $5000$ for $\Delta{\rm m}^2_{31}<0$ \\ 
%           & & & & & &  \\ % \hline
%\end{tabular}
%\caption{Different possible scenarios in the joint analysis} 
%\label{tab:scenariotable}
%\end{center}
%\end{table}

\section{Results of the simulation}
\label{sec3}
\subsection{Strategy}

Let us first describe briefly the strategy of our calculation.  There
are 6 measurable oscillation parameters in the 3-flavor neutrino
scheme. There are two mass scales; one is the atmospheric scale and
the other the solar scale.  Existing oscillation experiments have
determined that the two mass scales are widely separated and therefore
sensitive to different $L/E_\nu$ regions.  The LBL and VLBL we are
considering are affected by the solar scale only at next--to--leading
order, but can be strongly affected by the unknown parameters
$\delta_{CP}$ and $\theta_{13}$.  Hence we will take $\Delta{\rm
m}^2_{21}$ and $\theta_{12}$ to be the values determined in solar
experiments, as given in Eq. (\ref{input12}). This leaves 4 parameters
to be determined, but $|\Delta{\rm m}^2_{31}|$ and $\theta_{23}$ are
known already to a fair degree of accuracy from atmospheric neutrino
experiments.  Therefore obtaining $\theta_{13}$, $\delta_{\rm CP}$,
and the sign of $\Delta{\rm m}^2_{31}$ is the main goal of our
calculation.

The current and soon to be online LBL experiments will determine
$|\Delta{\rm m}^2_{31}|$ and $\sin^2(2\theta_{23})$ to a better
accuracy; $\delta m^2_{21}$ and $\sin^2(2\theta_{12})$ will be
measured more precisely by KamLAND~\cite{kamland,vbkam}. The first
task of the scenarios we propose is to determine $|\Delta{\rm
m}^2_{31}|$ and $\sin^2(2\theta_{23})$ more accurately, using the
$\nu_\mu$ survival probability. This will allow us to have as small an
uncertainty as possible in the determination of the parameters
$\theta_{13}$ and $\delta_{CP}$.  Let us define $N_{\alpha}(L)$ as the
number of charge--current events involving the $\alpha$ charged lepton
at the baseline $L$. We assume that $N_\mu(300)$, which depends on the
$\nu_\mu$ survival probability, is used to determine
$|\Delta m^2_{31}|$ and $\theta_{23}$ with as small an error as
possible. Then in the various scenarios $N_\mu (300)$ cannot
be used to determine $\theta_{13}$ and $\delta_{CP}$.~\footnote{When
new runs at $L$=300 km with better statistics are made, the improved
$N_\mu (300)$ will be used to update the values of $|\Delta m^2_{31}|$
and $\theta_{23}$.}
 
\subsection{Scenario I ($\nu_\mu$ beam only)}

\begin{figure}[htb]
\vspace*{-1cm}
\hspace*{-2cm}
\includegraphics[height=20cm,width=17cm,angle =0]{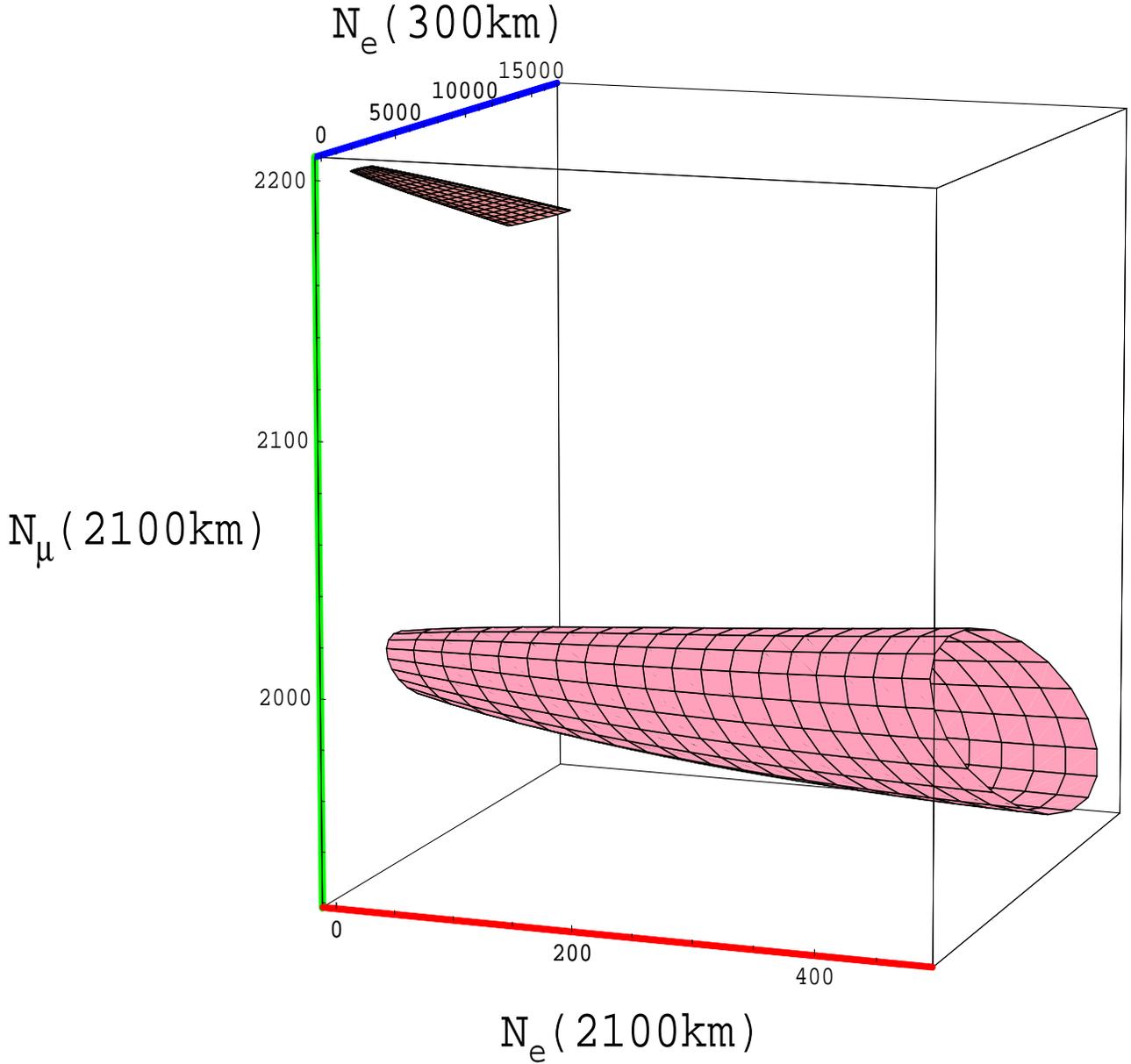}
\vspace*{-1cm}
\caption[]{  Three-dimensional surface in the events space
$N_e(300)-N_e(2100)-N_{\mu}(2100)$ in Scenario II with $CP$ phase $\delta_{CP}$
varying from $0$ to $2\pi$ and $\sin^2(2\theta_{13})$ from $0.01$ to $0.1$.
The lower (upper) one is for $\Delta m^2_{32}>0$ ($<0$). The surface in
Scenario I is obatined by scaling  $N_e(300)$ axis by a factor $\frac{0.77
({\rm MW})\times 5 ({\rm year}) \times 22.5 ({\rm kt})} {4 ({\rm MW})\times
2 ({\rm year}) \times 450 ({\rm kt})}\simeq \frac{1}{42}$.} 
\label{fig1}
\end{figure}

\begin{figure}[htb]
\vspace*{-1cm}
\hspace*{-2cm}
\includegraphics[height=20cm,width=17cm,angle =0]{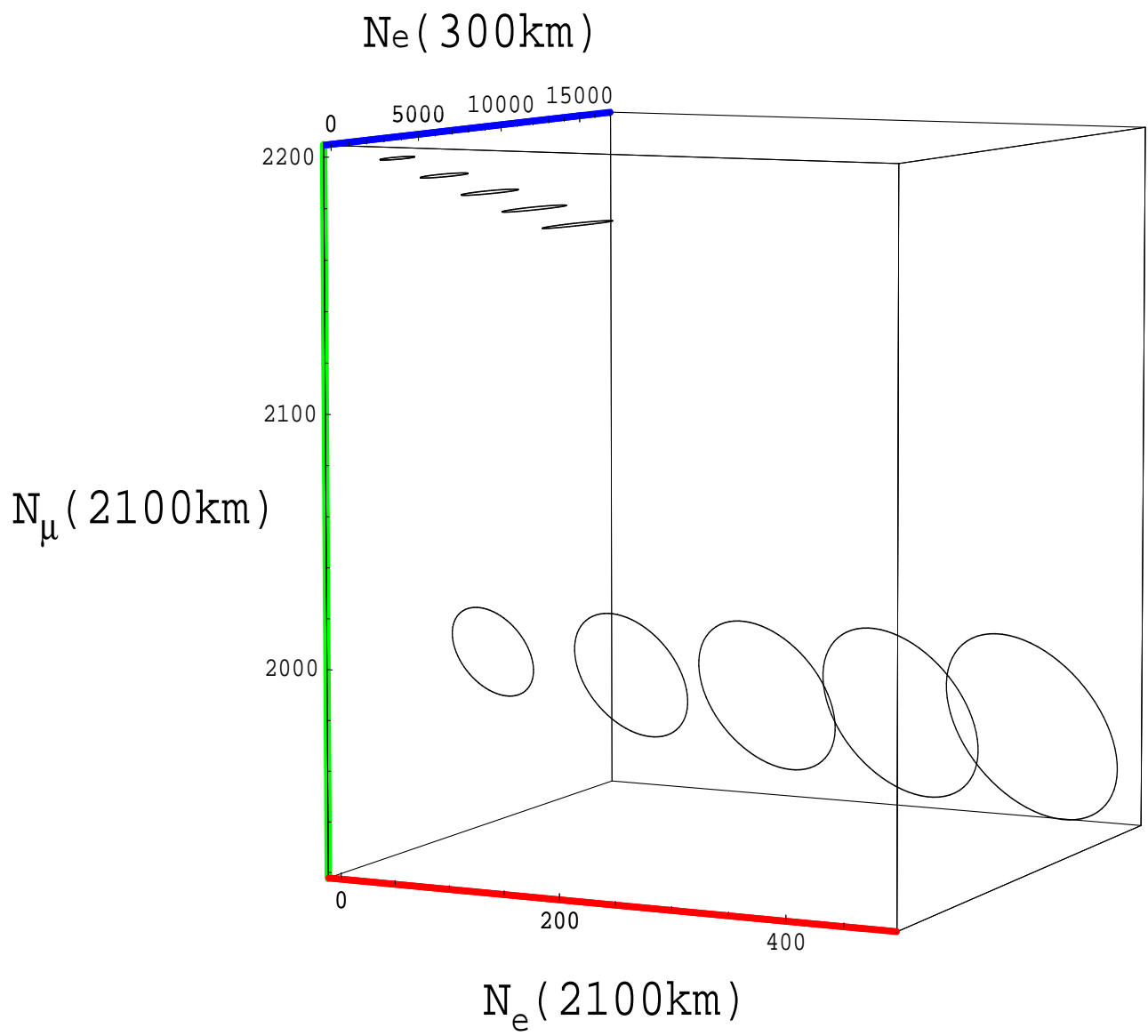}
\vspace*{-1cm}
\caption[]{ Same as Fig. \ref{fig1}, but for fixed
$\sin^2(2\theta_{13})=0.02$, $0.04$, $0.06$, $0.08$ and $0.1$ for the elipses
from left to right. }  
\label{fig2}
\end{figure}

In this scenario we assume that only the $\nu_{\mu}$ beam is employed
to run at $L$=300 km and 2100 km. 
Since $N_\mu (300)$ has already been used to determine $|\Delta m^2_{31}|$
and $\sin^2(2\theta_{23})$, we are left with 
three types of independent measurements for the determination of
$\theta_{13}$, $\delta_{\rm CP}$ and the sign of $\Delta m^2_{31}$: $N_e(300)$,
$N_e(2100)$ and $N_{\mu}(2100)$.  The measurements of these three types of
events form a surface in a three-dimensional space when $\theta_{13}$ and
$\delta_{CP}$ are varied in their allowed ranges. The angle $\theta_{13}$ is
constrained by the CHOOZ reactor experiment~\cite{CHOOZ} and the $CP$ phase
is completely unconstrained.  Therefore, we take their ranges to be:
$\sin^2(2\theta_{13})=(0, 0.1)$ and $\delta_{\rm CP} = (0,2\pi$).  Such
three--dimensional surfaces, which are tube--like, are displayed in
Fig.~\ref{fig1}. The upper and lower surfaces are for negative and
positive $\Delta{\rm m}^2_{31}$, respectively.  The closed curves around the
axes of the tubes are traced out by varying $\delta_{\rm CP}$ from $0$ to
$2\pi$, while the lines running parallel to the axes of the tubes are
determined by varying $\sin^2(2\theta_{13})$ from $0.01$ to $0.1$. For fixed
$\sin^2(2\theta_{13})$ values, we then obtain the ellipses in
Fig.~\ref{fig2}.  

\begin{figure}[htb]
\vspace*{-1cm}
\hspace*{-2cm}
\includegraphics[height=20cm,width=17cm,angle =0]{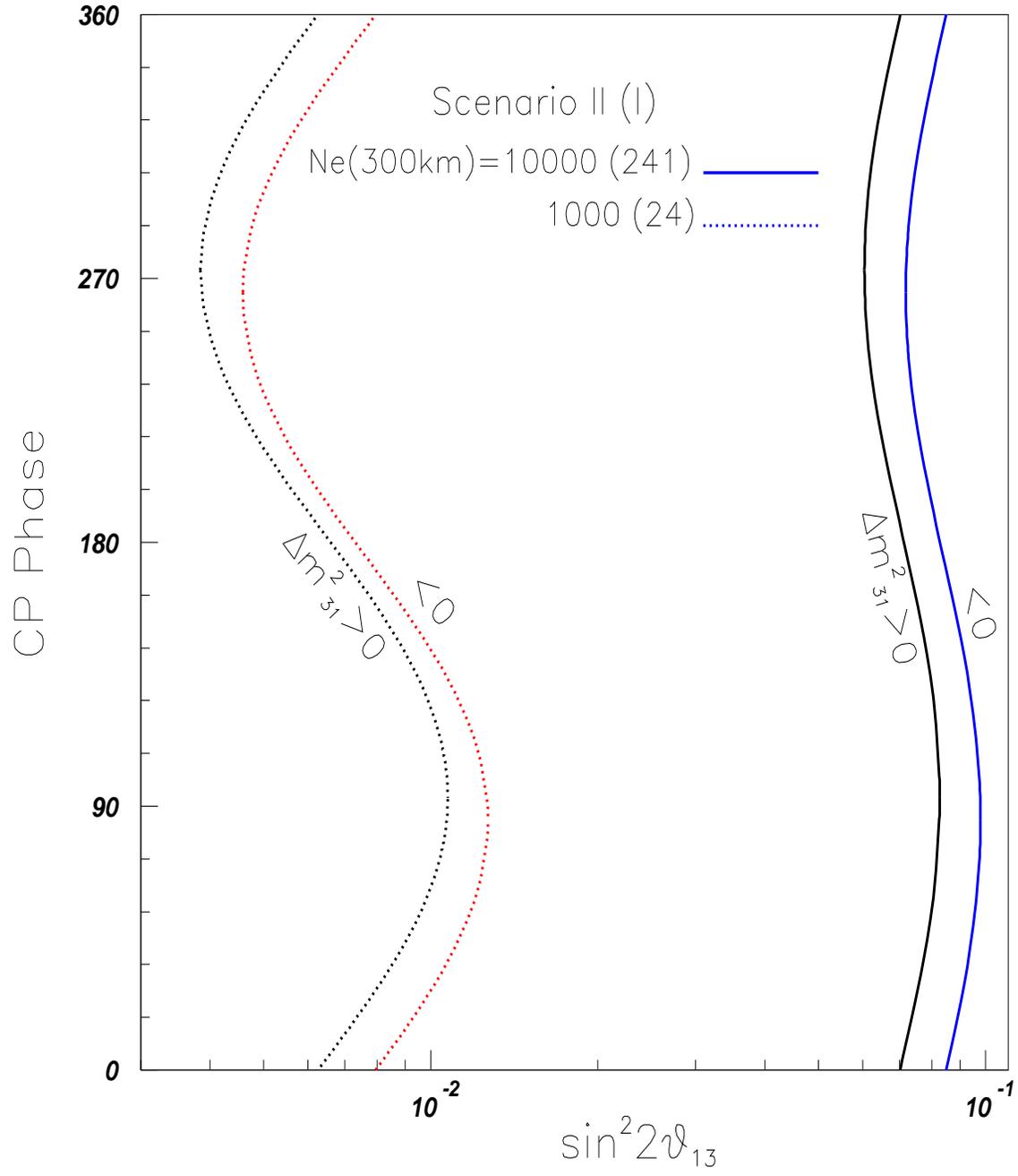}
\vspace*{1cm}
\caption[]{ $CP$ phase $\delta_{CP}$ (in degrees) versus $\sin^2(2\theta_{13})$
for fixed $N_e(300)$.}
\label{fig3}
\end{figure}

When $N_e(300)$, the number of the $\nu_e$ appearance events at 300 km, is
measured, it determines a closed curve which is obtained from the 
three-dimensional surface by a cut at a given value on the $N_e(300)$ axis.  
The value of $N_e(300)$ does not determine $\theta_{13}$ directly since
$\delta$ is unknown; for each of the closed curves we obtain a definite
relation between $\delta$ and $\sin^2(2\theta_{13})$ when the sign of 
$\Delta m^2_{31}$ is given.  We show in Fig.~\ref{fig3} two sets 
of such relations for each $\Delta{\rm m}^2_{31}$ sign.  As shown, we choose
two extreme values of $N_e(300)$, each of which leads to a range of values 
for $\sin^2(2\theta_{13})$, depending on the sign of 
$\Delta{\rm m}^2_{31}$.  For positive $\Delta{\rm m}^2_{31}$, the larger 
$N_e(300)$ curve limits $\sin^2(2\theta_{13})$ to the range 
($0.06,0.1$), while the smaller one corresponds to 
$\sin^2(2\theta_{13})$ lying in the range ($0.006,0.01$).  Similarly,
the ranges of the values of $\sin^2(2\theta_{13})$ for negative
$\Delta{\rm m}^2_{31}$ can be read off from Fig.~\ref{fig3}.

\begin{figure}[htb]
\vspace*{-1cm}
\hspace*{-2cm}
\includegraphics[height=20cm,width=17cm,angle =0]{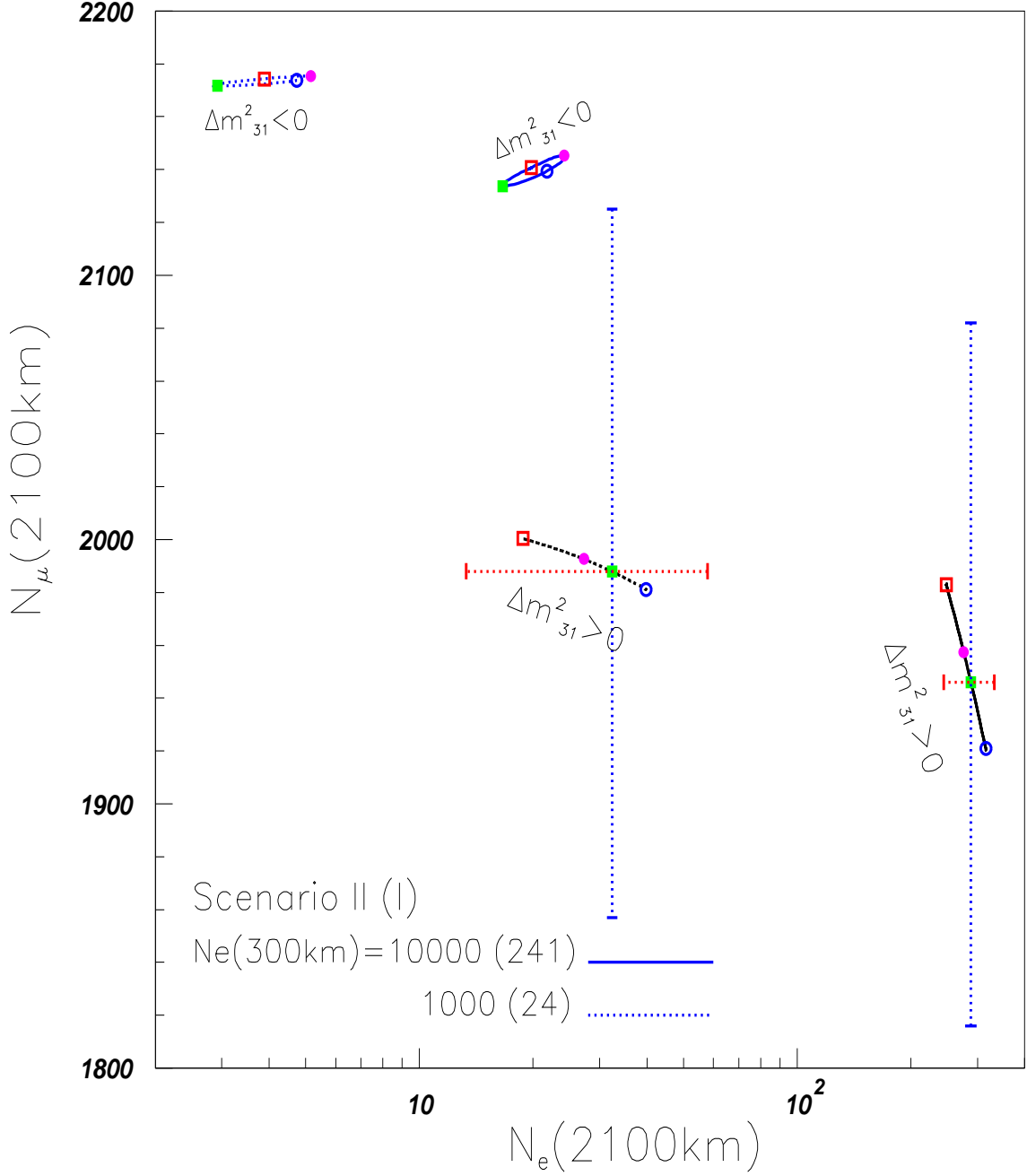}
\vspace*{-1cm}
\caption[]{ $N_{\mu}(2100)$ versus  $N_e(2100)$ for fixed $N_e(300)$. 
            The  open square,  open circle,  filled square and  filled circle denote 
            $\delta_{CP} = 0$, $\pi$, $\pi/2$ and  $3\pi/2$, respectively. 3$\sigma$ error                         bars at some points are also plotted.}
\label{fig4}
\end{figure}

We plot in Fig.~\ref{fig4} the two-dimensional curves with fixed $N_e(300)$.
Note that the scale of the horizontal axis is logarithmic.  If the scale was
linear, the curves would be ellipses. An open square indicates the point on
a curve with $\delta_{\rm CP}$=0$^\circ$, solid square 90$^\circ$, open circle
180$^\circ$, and solid circle 270$^\circ$. We also show a
representative 3$\sigma$ error bar for each curve. The assignment of
statistical and systematic errors has been discussed in the preceding section.
The total error is dominated mostly by the statistical error.  One sees that
although the sign of $\Delta m^2_{31}$ can be determined at the 3$\sigma$
level, there is no sensitivity to the value of the $CP$ phase. In particular,
the error in the $N_\mu(2100)$ channel is very large in comparison
with the range of variation in the number of events when $\delta_{\rm CP}$
varies; we will encounter similar situation in the next scenario.

\subsection{Scenario II ($\nu$ and $\bar\nu$ beams)}

\begin{figure}[htb]
\vspace*{-1cm}
\hspace*{-2cm}
\includegraphics[height=20cm,width=17cm,angle =0]{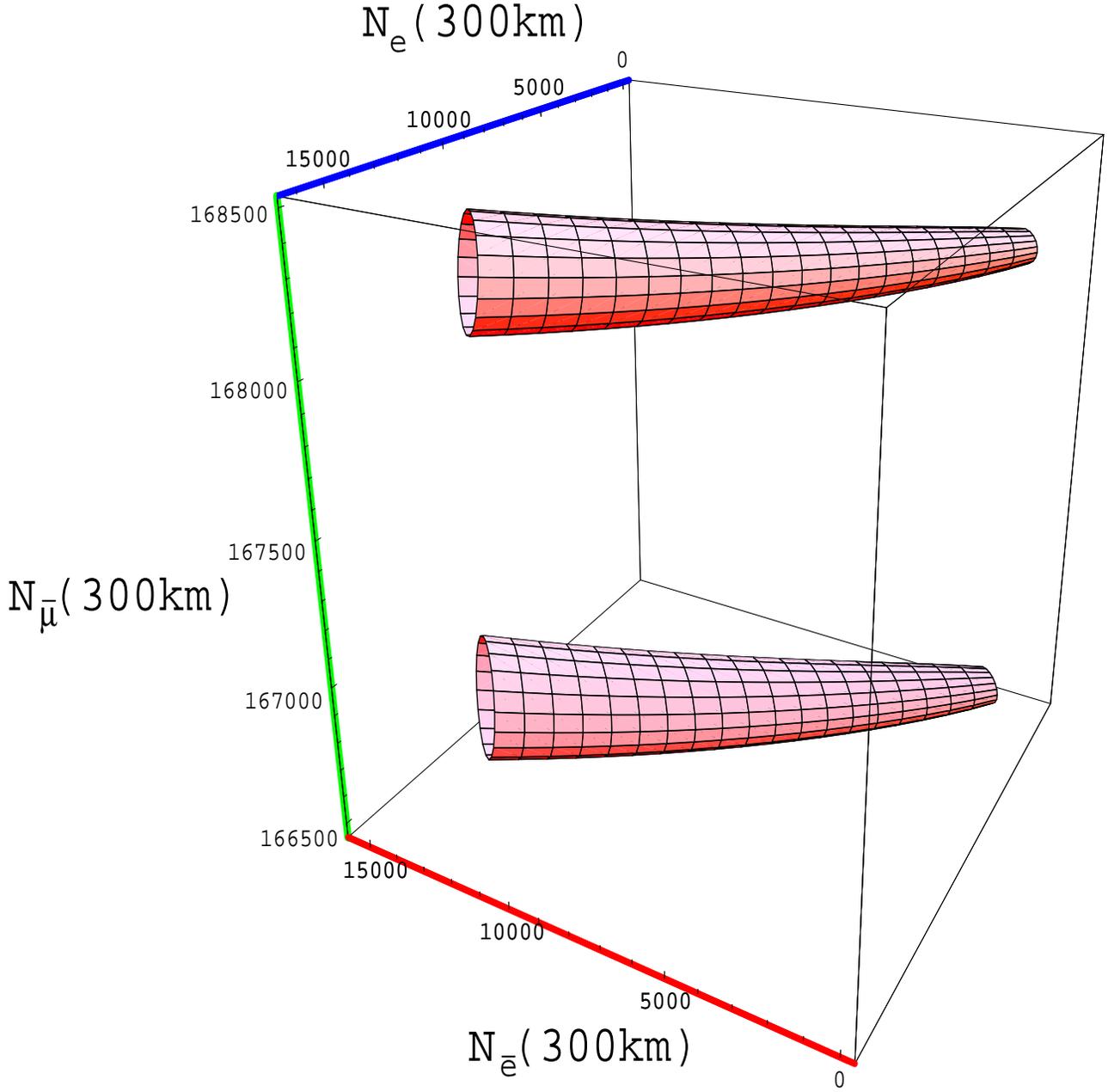}
\vspace*{-1cm}
\caption[]{ Three-dimensional surface in the events space $N_e(300)-N_{\bar e}(300)-N_{\bar \mu}(300)$
            in Scenario II with  $CP$ phase $\delta_{CP}$ varying from $0$ to $2\pi$ and  $\sin^2(2\theta_{13})$ 
            from $0.01$ to $0.1$.  The upper (lower) one is for $\Delta m^2_{32}>0$ ($<0$). }
\label{fig5}
\end{figure}

\begin{figure}[htb]
\vspace*{-1cm}
\hspace*{-2cm}
\includegraphics[height=20cm,width=17cm,angle =0]{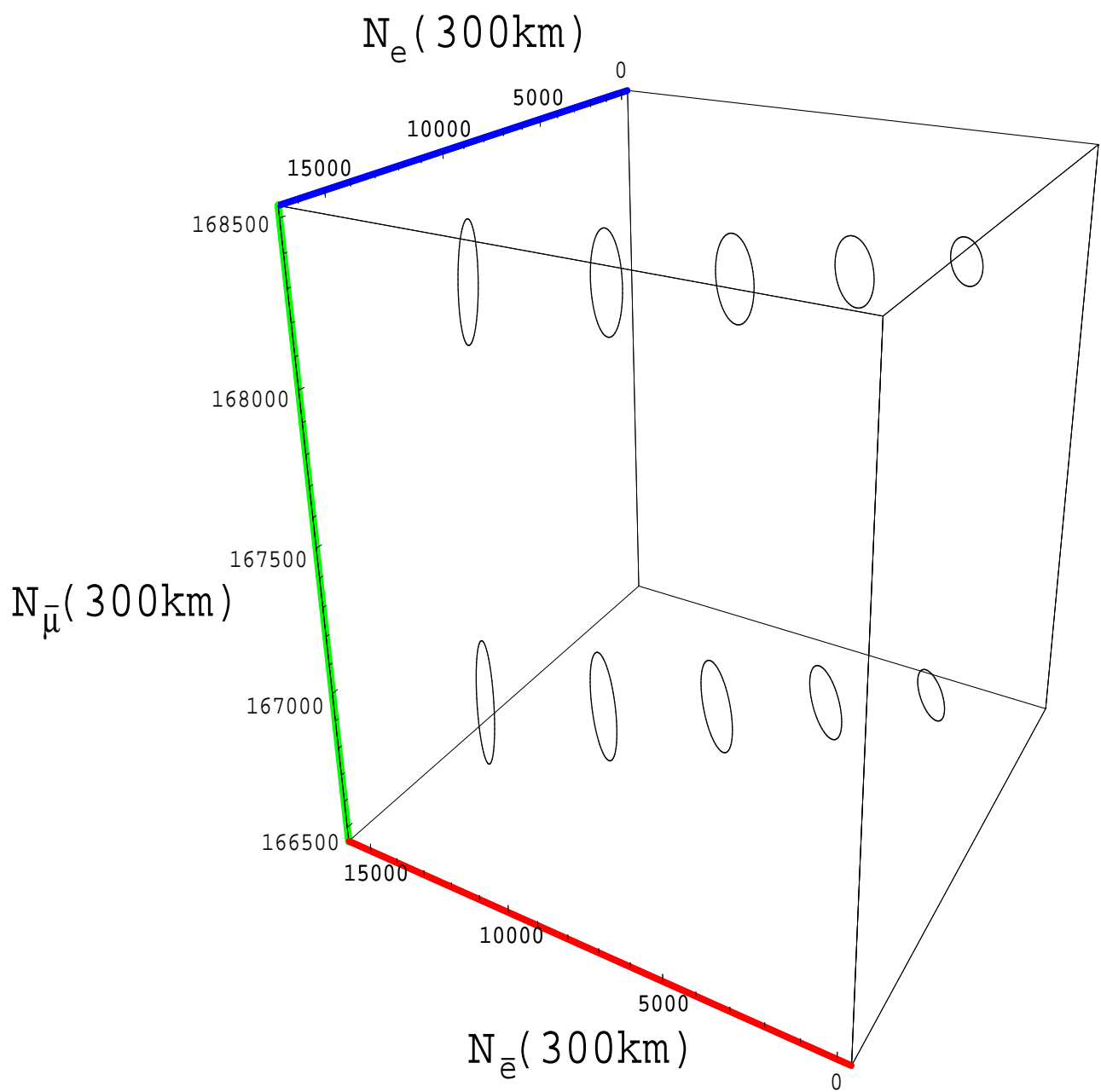}
\vspace*{-1cm}
\caption[]{ Same as Fig. \ref{fig5}, but for fixed 
 $\sin^2(2\theta_{13})=0.02$, $0.04$, $0.06$, $0.08$, $0.1$ for the elipses from right to left. }  
\label{fig6}
\end{figure}

\begin{figure}[htb]
\vspace*{-1cm}
\hspace*{-2cm}
\includegraphics[height=20cm,width=17cm,angle =0]{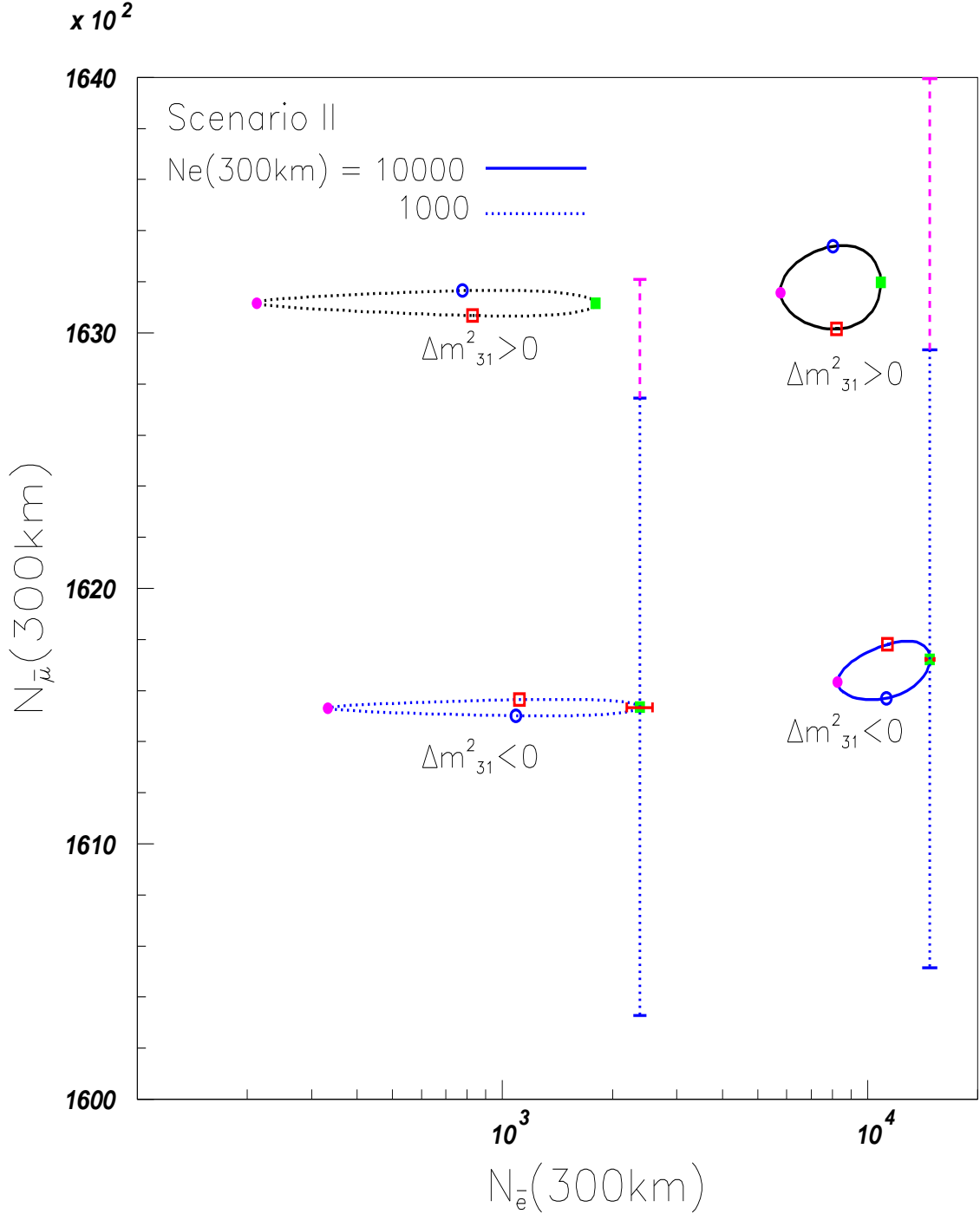}
\vspace*{-1cm}
\caption[]{ $N_{\bar \mu}(300)$ versus  $N_{\bar e}(300)$ for fixed $N_e(300)$ in Scenario II.
            The  open square,  open circle,  filled square and  filled circle denote 
            $\delta_{CP} = 0$, $\pi$, $\pi/2$ and  $3\pi/2$, respectively.
            3$\sigma$ error bars at some points are also plotted. 
            The dashed lines denote errors caused by $\theta_{23}$ uncertainty. 
            Note that only the upper error bar in  $N_{\bar \mu}$ changed since we
             take $\theta_{23}= \pi/4$ and any deviation from this always moves the
             $N_{\bar \mu}$ result in the same direction; this is an artifact of choosing
             maximal mixing as our starting point. 
 }
\label{fig7}
\end{figure}

\begin{figure}[htb]
\vspace*{-1cm}
\hspace*{-2cm}
\includegraphics[height=20cm,width=17cm,angle =0]{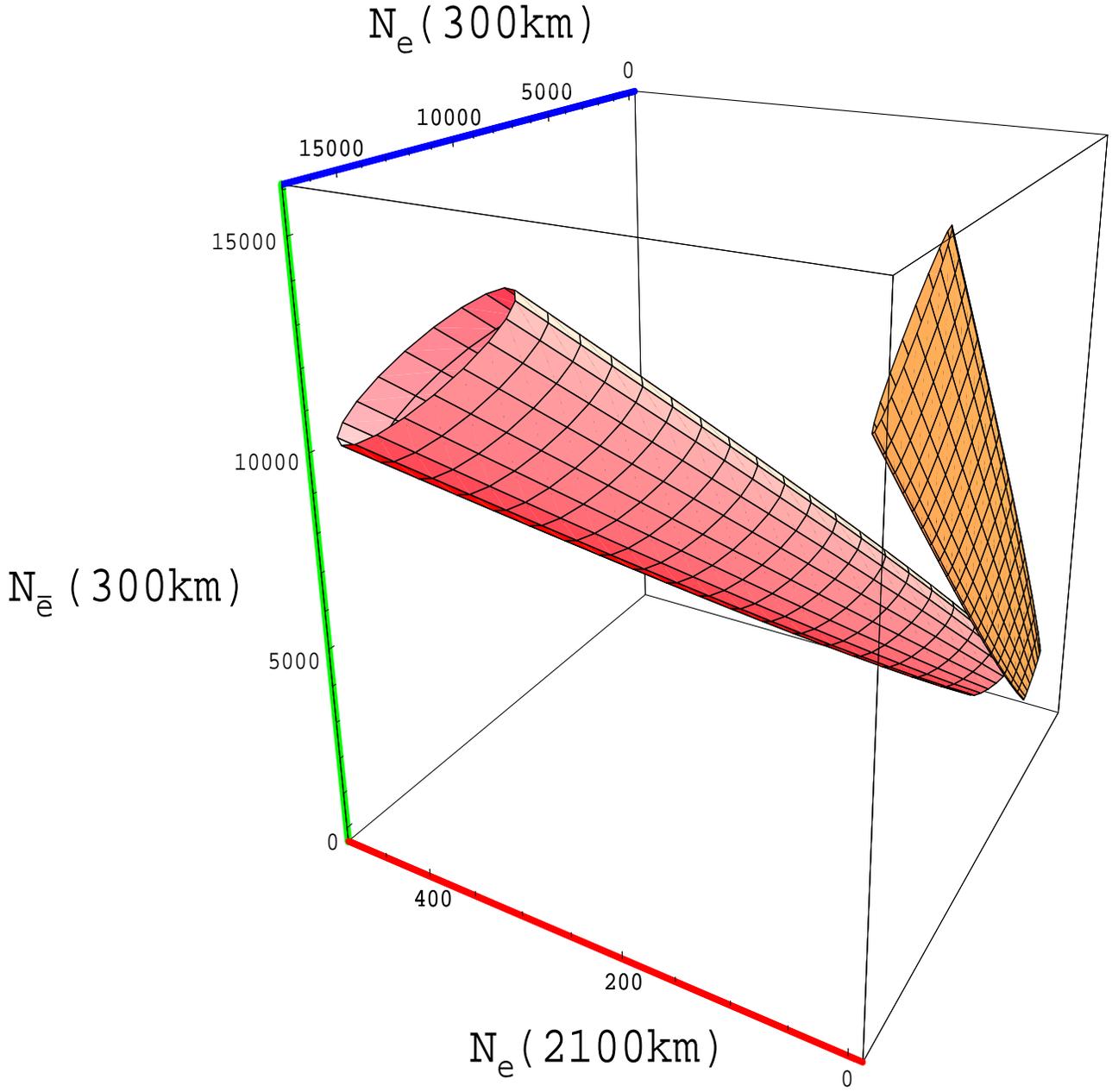}
\vspace*{-1cm}
\caption[]{ Three-dimensional surface in the events space 
$N_e(300)-N_e(2100)-N_{\bar e}(300)$
            in Scenario II with  $CP$ phase $\delta_{CP}$ varying from $0$ to $2\pi$ and $\sin^2(2\theta_{13})$ from $0.01$ to $0.1$. 
            The right (left) one is for $\Delta m^2_{32}>0$ ($<0$).  }  
\label{fig8}
\end{figure}

\begin{figure}[htb]
\vspace*{-1cm}
\hspace*{-2cm}
\includegraphics[height=20cm,width=17cm,angle =0]{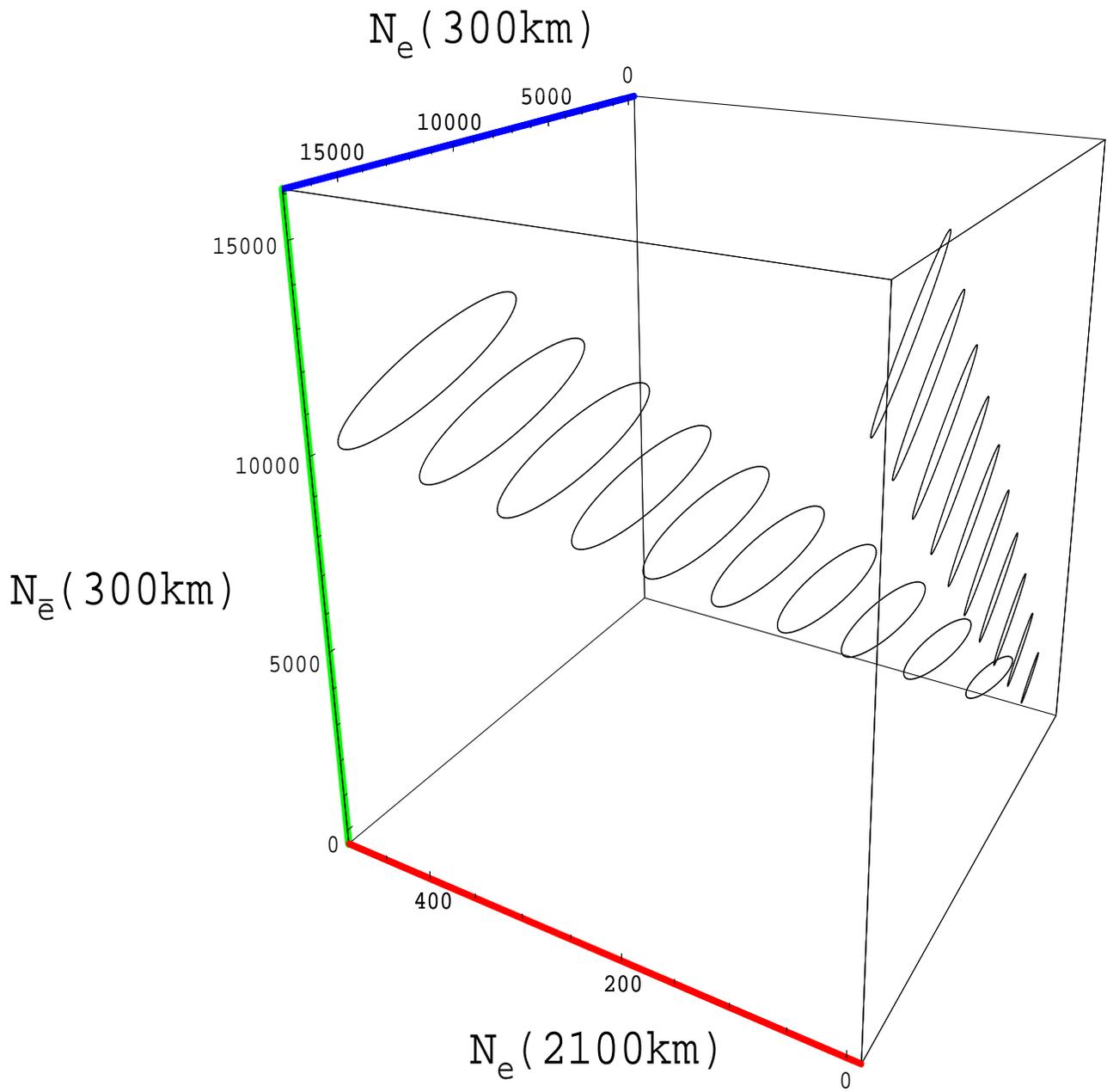}
\vspace*{-1cm}
\caption[]{ Same as Fig. \ref{fig8}, but for fixed $\sin^2(2\theta_{13})=0.01$, $0.02$, 
            $0.03$, $\cdot \cdot \cdot$, $0.1$ for the elipses from the lower-right corner 
            to the upper-left corner.
 }  
\label{fig9}
\end{figure}

\begin{figure}[htb]
\vspace*{-1cm}
\hspace*{-2cm}
\includegraphics[height=20cm,width=17cm,angle =0]{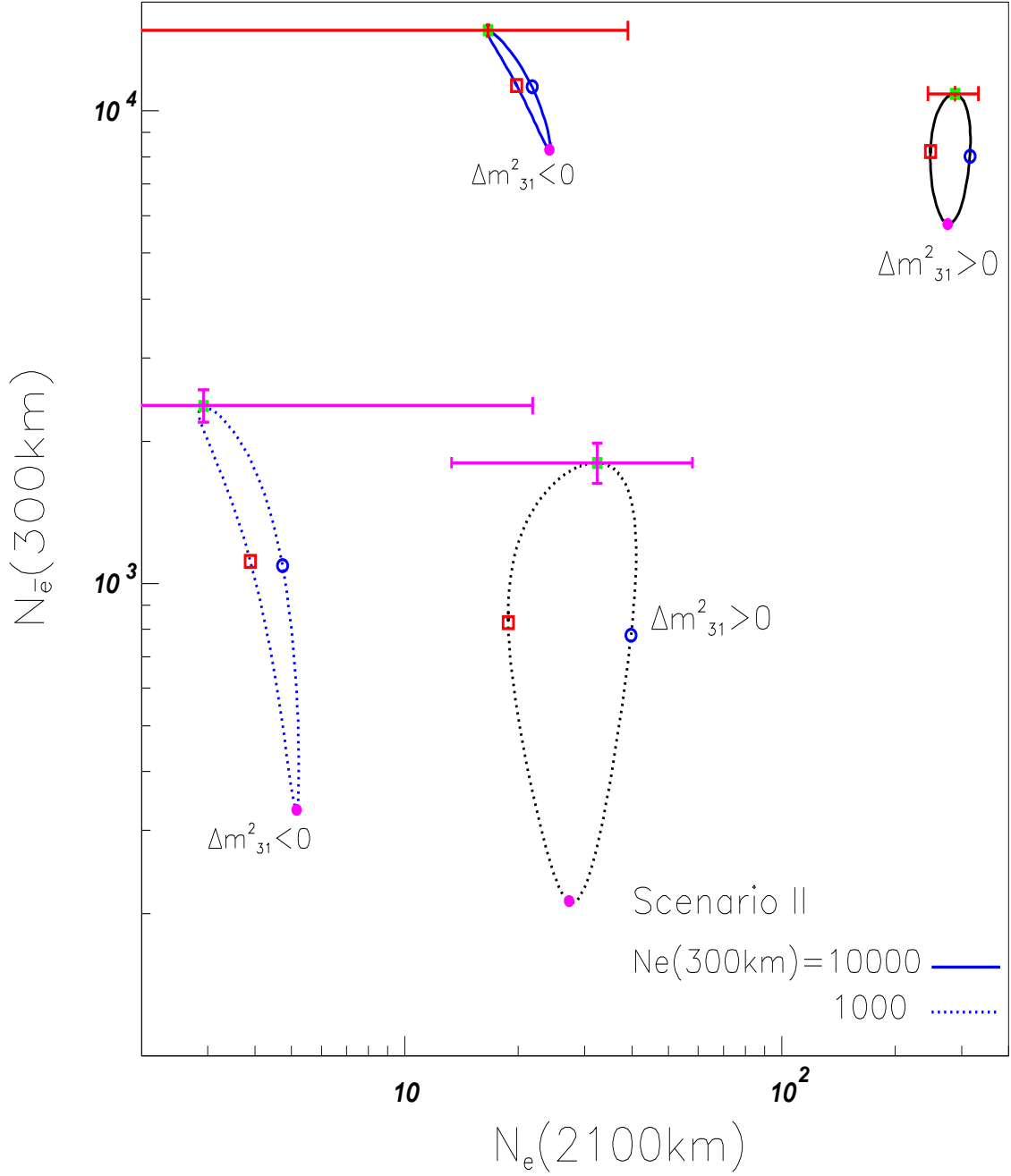}
\vspace*{-1cm}
\caption[]{ $N_{\bar e}(300)$ versus  $N_e(2100)$ for fixed $N_e(300)$  in Scenario II. 
            The  open square,  open circle,  filled square and  filled circle denote 
            $\delta_{CP} = 0$, $\pi$, $\pi/2$ and  $3\pi/2$, respectively.
            3$\sigma$ error bars at some points are also plotted.
 }
\label{fig10}
\end{figure}

By including the $\bar \nu_{\mu}$ beam aimed at the detector at
300~km, we have two more types of events, i.e., $N_{\bar e}(300)$
and $N_{\bar \mu}(300)$.  So, in addition to the three-dimensional surface
in the $N_e(300)$-$N_e(2100)$-$N_{\mu}(2100)$ space shown earlier in
Fig.~\ref{fig1}, we also have surfaces in the spaces
$N_e(300)$-$N_{\bar e}(300)$-$N_{\bar \mu}(300)$ and
$N_e(300)$-$N_{\bar e}(300)$-$N_e(2100)$, as shown in
Figs.~\ref{fig5} and \ref{fig8}, respectively.  With several
fixed values of $\sin^2(2\theta_{13})$ we obtain the curves shown in
Figs.~\ref{fig6} and \ref{fig9}.  We plot the
two-dimensional projections of fixed $N_e(300)$=1000 and 10000 in
Figs.~\ref{fig7} and \ref{fig10}.

We found in Scenario~I that the 2100 km data with just a $\nu_\mu$ source can
determine the sign of $\Delta m^2_{31}$ at 3$\sigma$, but cannot measure
the $CP$ phase (see Fig.~\ref{fig4}).  In contrast, as shown in
Fig.~\ref{fig7}, the 300 km data using both a $\nu$ and $\bar\nu$ source
can determine the $CP$ phase in the ranges ($\pi/2,3\pi/2$) or
($-\pi/2,\pi/2$), but cannot distinguish between the two ranges since the
measurement is only sensitive to $\sin\delta$.
Furthermore, unless $\delta_{CP}$ is close to $\pi/2$ or $3\pi/2$ the sign
of $\Delta m^2_{31}$ cannot be determined once all of the experimental
errors, including the error in the determination of $\theta_{23}$, are
taken into account,  leaving a four-fold ambiguity.  The problem lies in the
fact that $N_{\bar\mu}(300)$ is used; as already noted in Scenario I,
survival data provide poor resolution to the $CP$ phase, and the matter
effect is small at the relatively short distance of 300~km. Six more
three-dimensional plots which will contain either $N_{\bar \mu}(300)$, or
$N_\mu(2100)$, or both, can be made, but they are not very useful in the
present analysis because they involve the survival data.

In order to obtain good resolution in the sign of the MSD and to
distinguish the two ranges of the $CP$ phase as discussed in the
preceding section, we have to use data of the electron flavor
only. Hence we need two experiments with different $L/E_\nu$ ratios
and one of them should be a VLBL for a good sensitivity to the matter
effect. This brings us to the combined analysis of $N_e(300)$,
$N_{\bar e}(300)$ and $N_e(2100)$, as shown in Fig.~\ref{fig10}.  The
sign of $\Delta m^2_{31}$ can be easily determined if
$\sin^2(2\theta_{13})$ is not too small and the $CP$ phase can be
measured with again the ambiguity between the two ranges
($\pi/2,3\pi/2$) and ($-\pi/2,\pi/2$), as in the case of
Fig.~\ref{fig7}. The problem lies in the fact that the resolution in
$N_e(2100)$ is poor due to the low number of events, while the
resolution of the 300 km $\bar{\nu}_e$ is excellent. So we have to
increase the statistics at 2100 km.  This takes us to Scenario III
below.

\subsection{Scenario III ($\nu_\mu$ and $\bar\nu_\mu$ beams with increased
statistics)}

\begin{figure}[htb]
\vspace*{-1cm}
\hspace*{-2cm}
\includegraphics[height=20cm,width=17cm,angle =0]{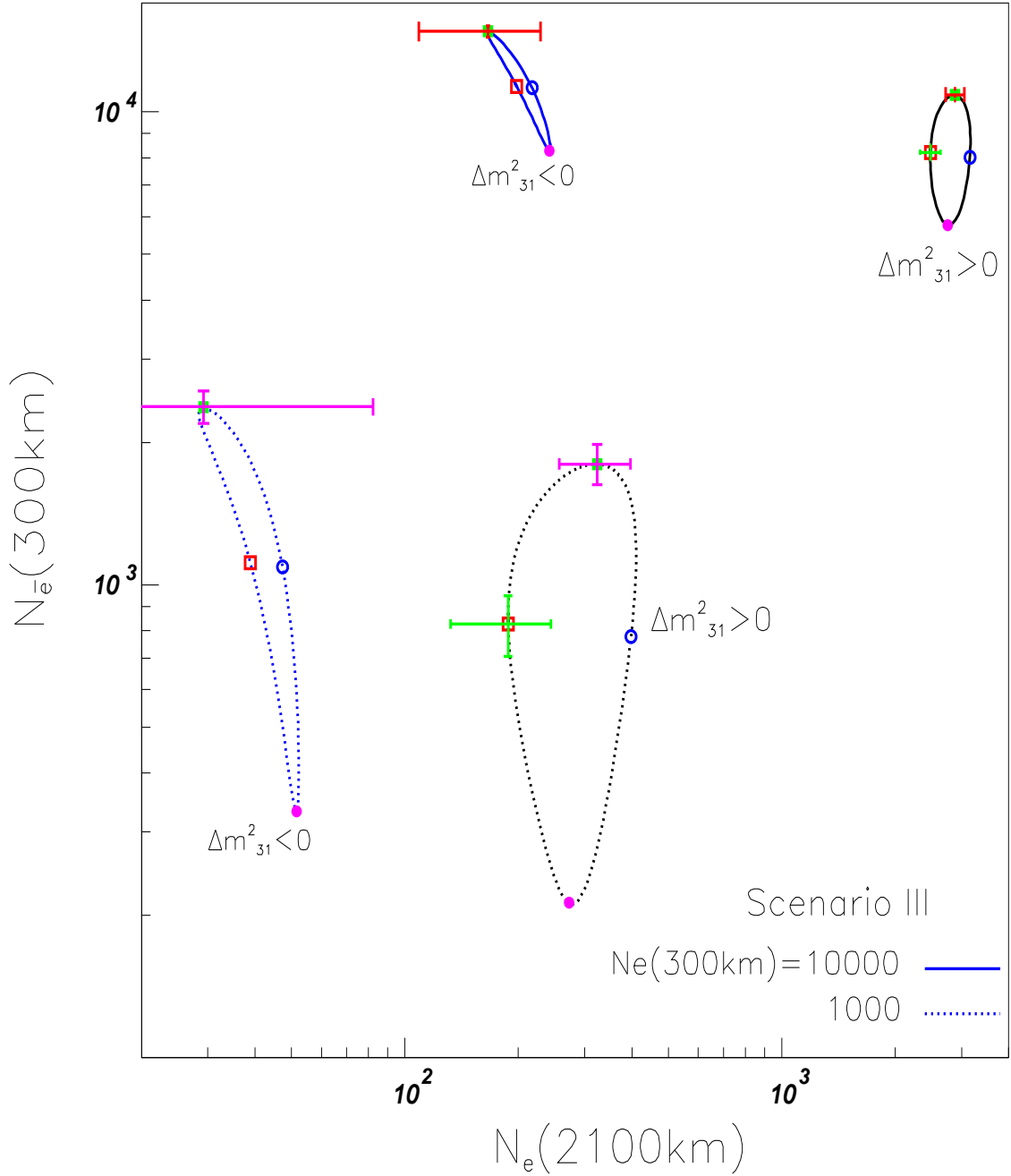}
\vspace*{-1cm}
\caption[]{ Same as Fig. \ref{fig10}, but for Scenario III.}  
\label{fig11}
\end{figure}

The situation of Scenario II can be improved if the statistics in
$N_e(2100)$ are significantly increased.  This can be achieved by
using a larger detector and/or running for a longer period of time for
the $N_e(2100)$ measurement.  For Scenario~III we set the detector
size times the running time at $L$=2100 km to be 10 times larger than
that of Scenario II, assuming that the number of events can be
straightforwardly scaled up with the detector size. The running at 300
km is the same as in Scenario~II.  The resultant two-dimensional plot
is shown Fig.~\ref{fig11}.  In this scenario, the sign of $\Delta m^2_{31}$
can be clearly determined at the 3$\sigma$ level, even for
$N_e(300)$=1000 which corresponds to a very small
$\sin^2(2\theta_{13})$ lying in the range ($0.006,0.01$), as indicated
by the dotted curves in Fig.~\ref{fig11}.

\begin{figure}[htb]
\vspace*{-1cm}
\hspace*{-2cm}
\includegraphics[height=20cm,width=17cm,angle =0]{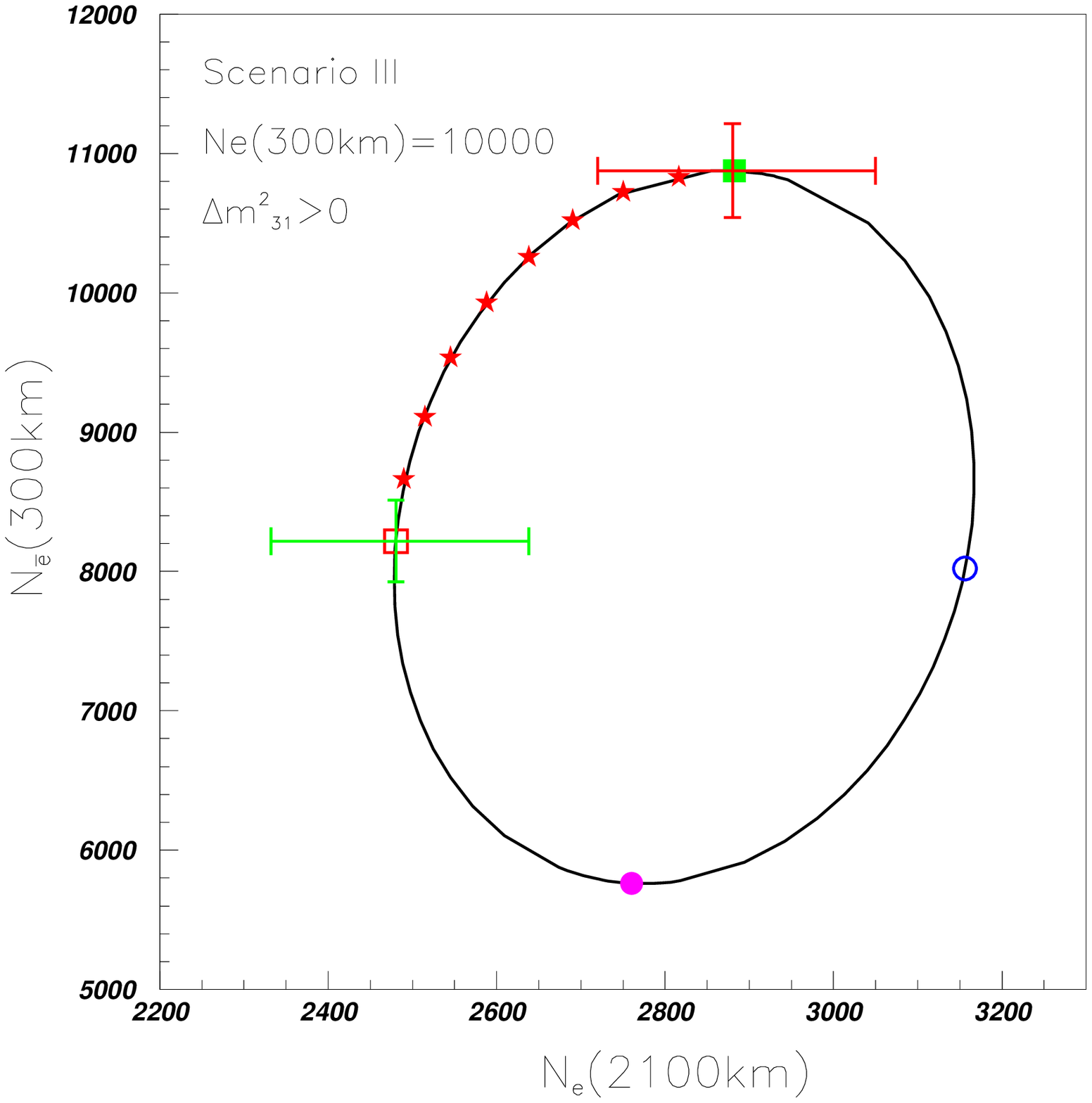}
\vspace*{-1cm}
\caption[]{ Same as Fig. \ref{fig11}, but for  $N_e(300)=10000$ and $\delta m^2_{31}>0$.
            The stars denote $10^{\circ}$ step of $CP$ phase from $0$ to $\pi/2$.
}  
\label{fig12}
\end{figure}

\begin{figure}[htb]
\vspace*{-1cm}
\hspace*{-2cm}
\includegraphics[height=20cm,width=17cm,angle =0]{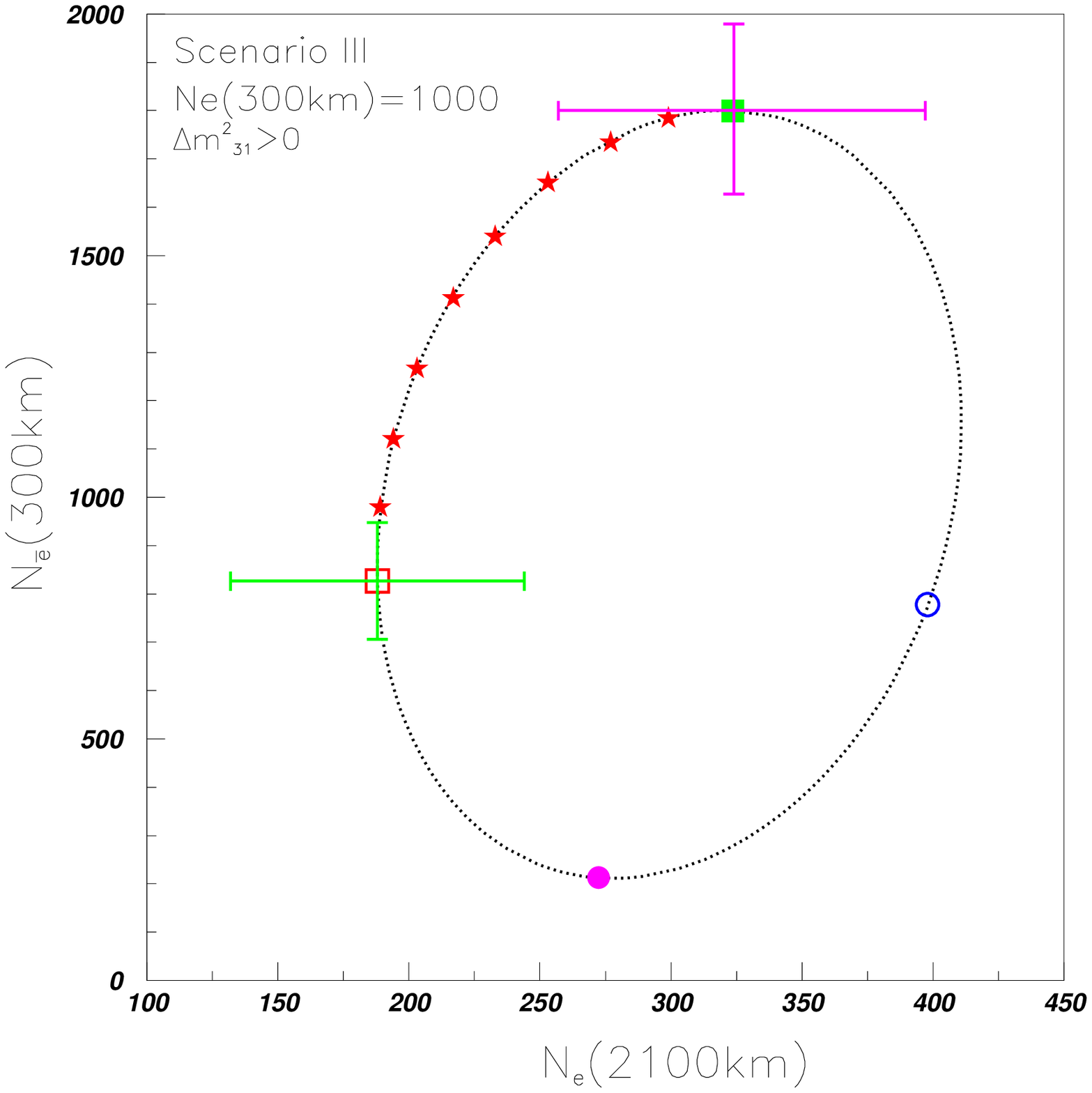}
\vspace*{-1cm}
\caption[]{ Same as Fig. \ref{fig11}, but for  $N_e(300)=1000$ and $\delta m^2_{31}>0$.
            The stars denote $10^{\circ}$ step of $CP$ phase from $0$ to $\pi/2$. 
}  
\label{fig13}
\end{figure}

If $\Delta m^2_{31}$ is positive, a reasonably accurate determination
of $\delta_{CP}$ can be made with no sgn($\Delta m^2_{31}$) or
$\theta_{13}$ ambiguity, and the $\bar \nu_{\mu}$ beam is not needed
at 2100 km even for very small $\sin^2(2\theta_{13})$ in the range
($0.006,0.01$). This is consistent with the results of
Ref.~\cite{bmw2001}, where it was found that a $\nu_\mu\to\nu_e$ and
$\bar\nu_\mu\to\bar\nu_e$ measurement at short distance and a
$\nu_\mu\to\nu_e$ measurement at a long distance could resolve
parameter ambiguities for $\sin^2(2\theta_{13}) > 0.005$.  To see the
sensitivity more clearly for positive $\Delta m^2_{31}$, we replot the
results in Figs.~\ref{fig12} and \ref{fig13} respectively
for $N_e(300)$=10000 and 1000. We see that for $N_e(300)=10000$, which
corresponds to larger $\sin^2(2\theta_{13})$ ($0.06 - 0.1$) as shown
in Fig.~\ref{fig3}, the $CP$ phase can be determined better than
10$^\circ$ at 3$\sigma$ for $\delta_{\rm CP}$ small or around
180$^\circ$.  The sensitivity deteriorates slowly when $\delta_{CP}$
moves away from $0^\circ$ or $180^\circ$, and the uncertainty becomes of
the order of 25$^\circ$ when $\delta_{\rm CP}$ is close to 90$^\circ$
or 270$^\circ$.

Even for $N_e(300)$=1000, which corresponds to very small
$\sin^2(2\theta_{13})$ in the range of ($0.006,0.01$), the measurement
of the $CP$ phase is still reasonably good.  It is interesting to note
that the sensitivity of the $CP$ measurement near $\delta_{\rm
CP}$=0$^\circ$ and 180$^\circ$ for $N_e(300) = 1000$ is comparable to
that of the much higher number of events of $N_e(300)$=10000.  Hence,
in this scenario, either case can establish whether or not $CP$ in the
lepton sector is violated if $\delta_{\rm CP}$ deviates by than
10$^\circ$ from the $CP$ conserving points of $\delta_{\rm
CP}$=0$^\circ$ or 180$^\circ$.

\begin{figure}[htb]
\vspace*{-1cm}
\hspace*{-2cm}
\includegraphics[height=20cm,width=17cm,angle =0]{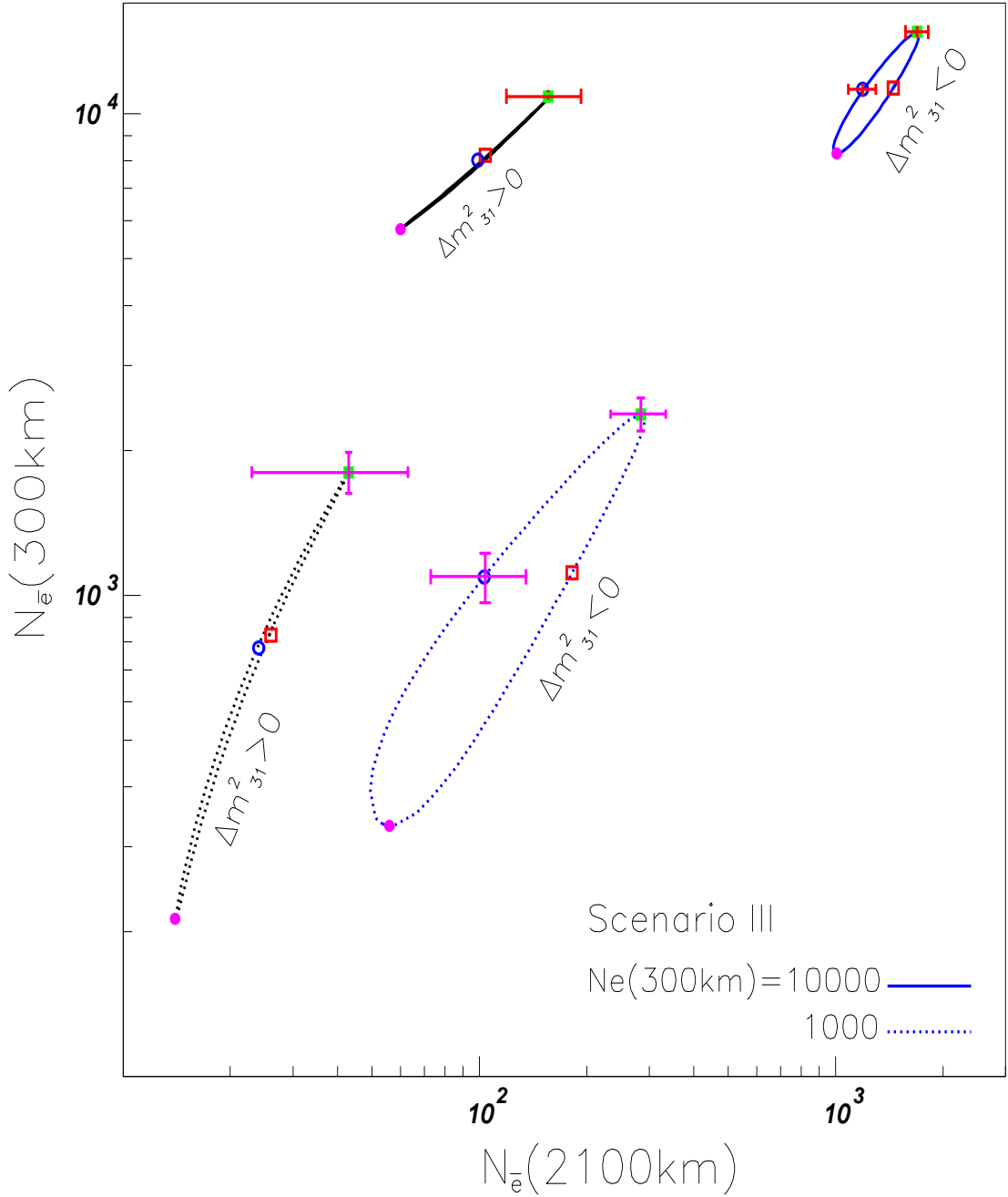}
\vspace*{-1cm}
\caption[]{ Same as Fig. \ref{fig11}, but for  $N_{\bar e}(300)$ versus  $N_{\bar e}(2100)$.
}  
\label{fig14}
\end{figure}

\begin{figure}[htb]
\vspace*{-1cm}
\hspace*{-2cm}
\includegraphics[height=20cm,width=17cm,angle =0]{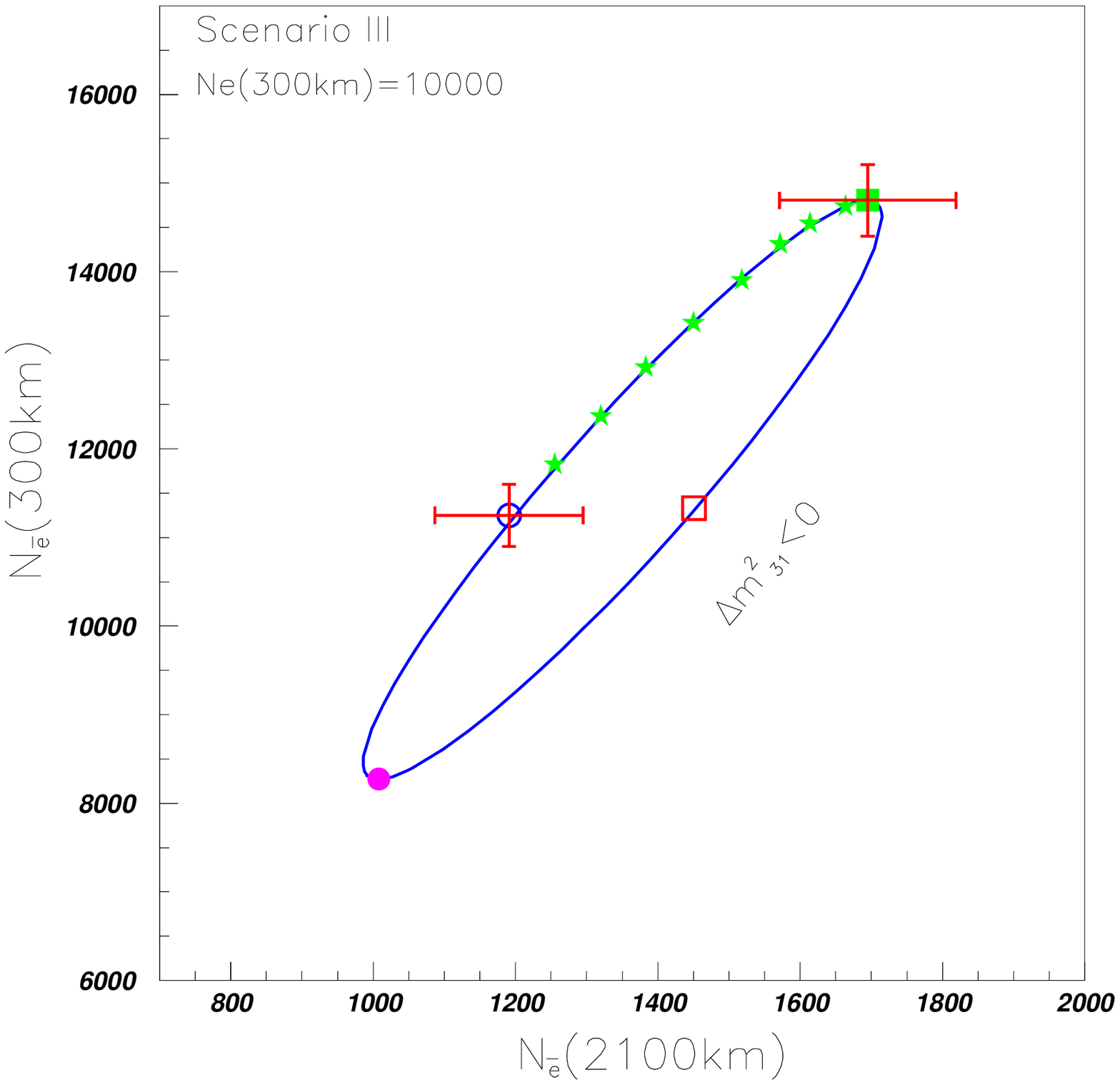}
\vspace*{-1cm}
\caption[]{ Same as Fig. \ref{fig14}, but for  $N_e(300)=10000$ and $\delta m^2_{31}<0$.
            The stars denote $10^{\circ}$ step of $CP$ phase from $\pi/2$ to $\pi$.
}  
\label{fig15}
\end{figure}

\begin{figure}[htb]
\vspace*{-1cm}
\hspace*{-2cm}
\includegraphics[height=20cm,width=17cm,angle =0]{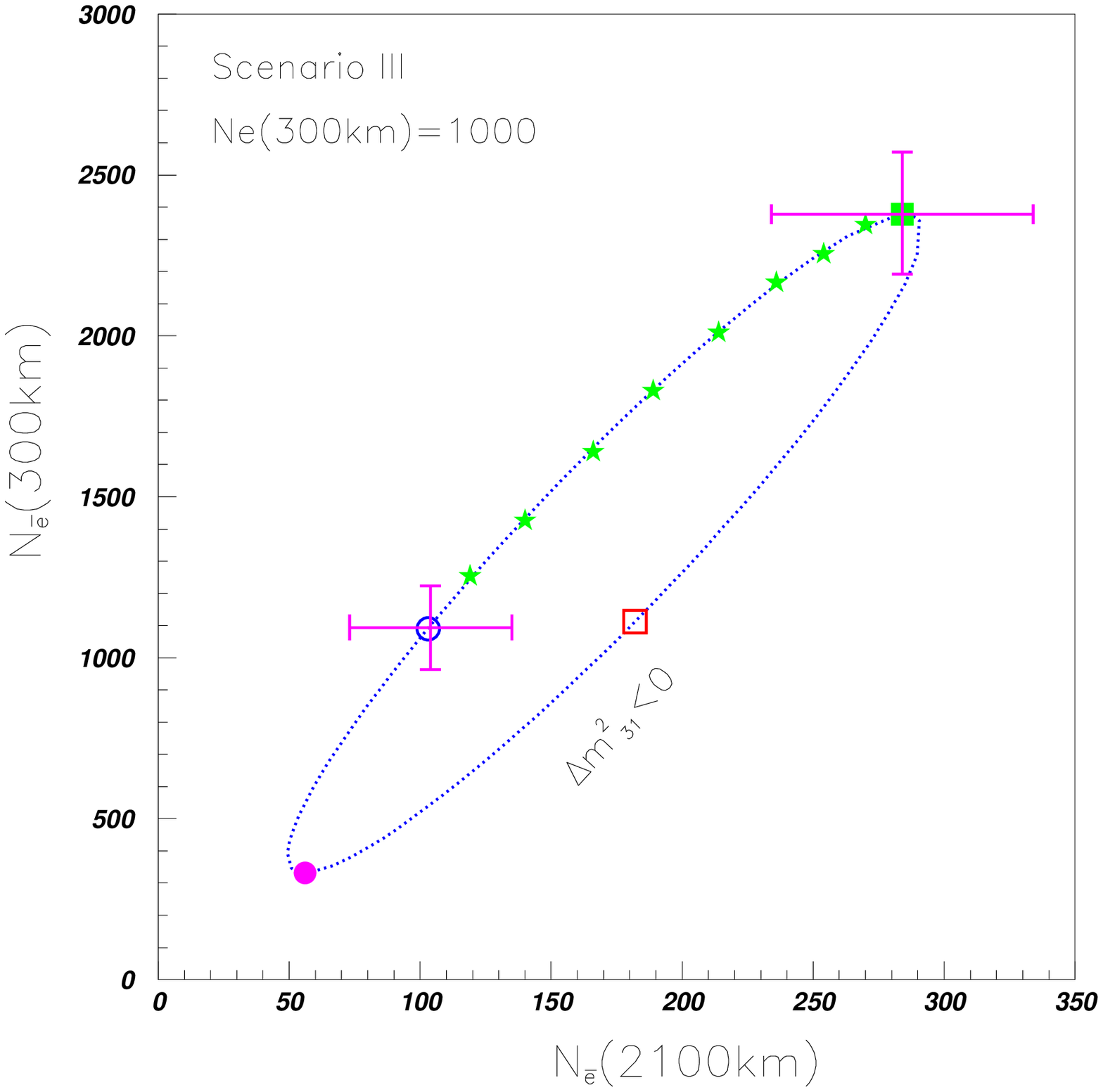}
\vspace*{-1cm}
\caption[]{ Same as Fig. \ref{fig14}, but for  $N_e(300)=1000$ and $\delta m^2_{31}<0$.
            The stars denote $10^{\circ}$ step of $CP$ phase from $\pi/2$ to $\pi$.
}  
\label{fig16}
\end{figure}

If $\Delta m^2_{31}$ is negative, the $\bar\nu_\mu \rightarrow
\bar\nu_e$ oscillation is the favorable channel to investigate.  Hence
once it is clear that $\Delta{\rm m}^2_{31}$ is negative (see the next
section for a detailed discussion), the $\bar \nu_{\mu}$ beam should
be delivered to 2100 km to run for 5000 kt-yr.  The results, which are
the counterparts to Fig.~\ref{fig11}, are shown in
Fig.~\ref{fig14}. With positron events at 2100 km the $CP$ phase
can be well measured. To see the sensitivity more clearly, we replot
the results in Figs.~\ref{fig15} and \ref{fig16} for
$N_e(300)$=10000 and 1000 respectively.  The accuracy of the
$\delta_{CP}$ measurement for $\Delta m^2_{31} < 0$ using the
$\bar\nu_\mu$ beam is about the same as that of the $\nu_\mu$ beam for
$\Delta{\rm m}^2_{31} > 0$, although the distinction between
$\delta_{CP}$ in the range ($-\pi/2,\pi/2$) and $\delta_{CP}$ in the
range ($\pi/2,3\pi/2$) is not as good for a $\bar\nu_\mu$ beam with
$\Delta m^2_{31} < 0$.

We have also done the analysis assuming a $\nu$ or $\bar\nu$ NBB of
peak energy 5, 6, or 8~GeV is delivered to the detector at 2100~km. We
found that for these cases the ellipses in Figs.~\ref{fig11} to
\ref{fig15} are much flatter than for the 4~GeV NBB, so that
they do not do as well in resolving the degeneracy in $\delta_{CP}$.

\section{Conclusion and discussion}

We conclude that with a superbeam, such as that delivered by HIPA, the
joint analysis at two baselines, of which one is an LBL at 300 km and the
other a VLBL at 2100 km, can determine the $\Delta m^2_{31}$ sign and
give a reasonably precise measurement of the $CP$ phase and
$\theta_{13}$.  To achieve this, both $\nu_\mu$ and $\bar\nu_\mu$
beams are needed for the LBL experiment. The survival events
$\nu_\mu\rightarrow\nu_\mu$ and
$\bar\nu_\mu\rightarrow\bar\nu_\mu$ are generally insensitive to
the matter and $CP$ effects.  

The initial HIPA $\nu_\mu$ beam with power 0.77 MW will run with
exposure 22.5$\times$5 kt-yr at a detector at $L$=300 km to obtain
both survival events $N_\mu(300)$ and appearance events
$N_e(300)$. The former is used to improve the determination of the
mixing angle $\theta_{23}$ and mass--squared difference $|\Delta
m^2_{31}|$, so as to reduce the uncertainty of these crucial input
parameters. The latter can show the existence of an appearance signal
for $\sin^2(2\theta_{13}) \ge 0.006$ and find a crude relation between
$\delta_{CP}$ and $\theta_{13}$ as shown in Fig.~\ref{fig3}.

A detailed determination of the oscillation parameters will require an
upgrade of the HIPA beam power to 4~MW. Using our studies in this paper
as a guide, we suggest as one possibility the following experimental
steps using the upgraded HIPA beam:
\vskip 1ex

%\begin{description}{\setlength\itemsep{-0.6ex}
%\item

{\bf Stage 1}: Deliver a 4~MW $\nu$ 2$^\circ$--OAB to a 450~kt detector
        at a distance of $L$=300~km for 2 years. The survival events
        $N_\mu(300)$ are used to determine more precisely the
        parameters $\theta_{23}$ and $|\Delta{m}^2_{31}|$. The
        appearance events $N_e(300)$ are used to refine the relation
        between $\delta_{CP}$ and $\theta_{13}$, as shown in
        Fig.~\ref{fig3}.

{\bf Stage 2}: A 4~MW $\nu$ NBB with peak energy around 4~GeV is
        delivered to a detector at $L$=2100 km, to run for
        100$\times$5 kt-yr.  The survival and appearance events
        $N_\mu(2100)$ and $N_e(2100)$ are used to determine the sign
        of $\Delta{m}^2_{31}$, with the most sensitivity coming from
        $N_e(2100)$ (see Fig.~\ref{fig4}).

{\bf Stage 3}: A 4~MW $\bar{\nu}_\mu$ 2$^\circ$--OAB is delivered to
        the 300 km baseline detector for 450$\times$6 kt-yr and
        $N_{\bar\mu}(300)$ and $N_{\bar e}(300)$ are obtained.  The
        data can only determine $\delta_{CP}$ and $\theta$ up to a
        2-fold degeneracy because of the poor separation between
        $\delta$ and $\pi - \delta$ in the $N_{\bar\mu}(300)$
        measurement, as demonstrated in Fig.~\ref{fig7}.

{\bf Stage 4}: A 4 MW $\nu_\mu$ ($\bar\nu_\mu$) NBB with peak energy
        around 4~GeV is delivered to the 2100 km baseline detector for
        1000$\times$5 kt-yr if $\Delta{m}^2_{31} > 0$
        ($\Delta{m}^2_{31} < 0$). Then at 3$\sigma$, the value of
        $\delta_{CP}$ can be determined to about $10^\circ$ for values
        close to $0^\circ$ or $180^\circ$, or to about $25^\circ$ for
        values close to $90^\circ$ or $270^\circ$. The distinction
        between $\delta_{CP}$ in the ranges $(-\pi/2,\pi/2)$ and
        $(\pi/2,3\pi/2)$ is better for $\Delta m^2_{31} > 0$ than for
        $\Delta m^2_{31} < 0$. The $\Delta{m}^2_{31} >0$ case is shown
        in Figs.~\ref{fig12} and \ref{fig13} and that of
        $\Delta{m}^2_{31} <0$ in Figs.~\ref{fig15} and
        \ref{fig16}.

%}\end{description}
\vskip 1ex

% Should $\Delta{\rm m}^2_{31}$ be negative, a 
%$\bar \nu_{\mu}$ beam would also be necessary at the VLBL.  The $\nu_\mu$ 
%survival events in the present discussion are only used initially for the
%determination of $\theta_{23}$ but are not useful for the $CP$ and
%$\theta_{13}$ measurements. It is obvious that the capability to distinguish
%$\mu$ from $e$ is critical in the future LBL experiments.

It is apparent from our calculation that in order to obtain enough
statistics to provide a reasonably precise measurement of
$\theta_{13}$ and $\delta_{CP}$ the total detector size and running
time have to be sufficiently large. We have not attempted a detailed
optimization; rather, we offer our calculation as an example for
illustration.  A search is still required to determine the optimal
conditions for the measurement.  Eventually uncertainties in the
Earth matter density along a given baseline as well as uncertainties
in the solar neutrino oscillation parameters $\theta_{12}$ and $\Delta
m^2_{21}$ must also be taken into account.

\section*{Acknowledgment}

We thank T. Kobayashi for providing neutrino beam profiles and K. Hagiwara 
for discussions. We also thank L.-Y. Shan and our colleagues of the H2B 
collaboration \cite{H2B} for discussions.  This work is supported in part by 
DOE Grant No. DE-FG02-G4ER40817.

\newpage

%\begin{table}[hbtp!]
\begin{center}
Table 1 Different possible scenarios in joint analyses
\vskip 3ex 
\begin{tabular}{|l|c|c|c|c|c|c|}  \hline   
           & \multicolumn{3}{c|}{ } & \multicolumn{3}{c|}{ } \\ 
           & \multicolumn{3}{c|}{L=300 km} & \multicolumn{3}{c|}{L=2100 km} \\ 
           & \multicolumn{3}{c|}{ } & \multicolumn{3}{c|}{ } \\  \cline{2-7}
           & & & & & & \\ 
           & beam & power & detector size $\times$  runing time  & beam & power
                                      & detector size $\times$ runing time \\  
           & (2$^\circ$--OAB)& (MW)  & (kt$\cdot$ year)  & (NBB)&(MW) & (kt$\cdot$year)       \\ 
           & & & & & &  \\ \hline \hline
           & & & & & & \\
Scenario I & $\nu_{\mu}$ &  $0.77$ & $22.5\times 5$ & $\nu_{\mu}$ & $4$  
                & $100\times 5$  \\  
           & & & & & & \\ \hline \hline
           & & & & & &  \\
Scenario II  & $\nu_{\mu}$&    $4$    & $450\times  2$    & $\nu_{\mu}$& $4$  
                  & $100\times 5$  \\ 
           & & & & & & \\ 
            & $\bar \nu_{\mu}$& $4$    & $450\times  6$       &            &      &  \\ 
           & & & & & &  \\ \hline \hline
           & & & & & &  \\
Scenario III & $\nu_{\mu}$    & $4$    & $450\times  2$  & $\nu_{\mu}$     & $4$ 
          & $5000$ for $\Delta{\rm m}^2_{31}>0$ \\ 
           & & & & & &  \\
          & $\bar \nu_{\mu}$& $4$    & $450\times  6$  & $\bar \nu_{\mu}$  & $4$ 
           & $5000$ for $\Delta{\rm m}^2_{31}<0$ \\ 
           & & & & & &  \\  \hline
\end{tabular}
%\caption{Different possible scenarios in the joint analysis} 
%\label{tab:scenariotable}
\end{center}
%\end{table}
\newpage
%%% mu(2100)-e(2100)-e(300) %%%%%%%%%%%%%%%%%%%%%%%%%%%%
%%%%%%%%%%%%%%%%%%%%%%%%%%%%%%%
%%%%%%%%%%%%%%%%%%%%%%%%%%%%%%%
%%%%%%%%%%%%%%%%%%%%%%%%%%%%%%%
%%%%% mub(300)-eb(300)-e(300) %%%%%%%%%%%%%%%%%%%%%%%%%%
%%%%%%%%%%%%%%%%%%%%%%%%%%%%%%%
%%%%%%%%%%%%%%%%%%%%%%%%%%%%%%
%%%%%% eb(300)-e(2100)-e(300) %%%%%%%%%%%%%%%%%%%%%%%%%
%%%%%%%%%%%%%%%%%%%%%%%%%%%%%%%
%%%%%%%%%%%%%%%%%%%%%%%%%%%%%%
%%%%%%%%%%%%%%%%%%%%%%%%%%%%%%
%%%%%%%%%%%%%%%%%%%%%%%%%%%%%%
%%%%%%%%%%%%%%%%%%%%%%%%%%%%%%
%%%%%%%%%%%%%%%%%%%%%%%%%%%%%%
%%%%%%%%%%%%%%%%%%%%%%%%%%%%%%
%%%%%%%%%%%%%%%%%%%%%%%%%%%%%%

\end{document}